\documentclass[reprint,
superscriptaddress,
amsmath,amssymb,aps,showkeys,showpacs,
twoside,final,secnumarabic,
nofootinbib,onecolumn]{revtex4-2}

\usepackage[paperwidth=205mm,paperheight=290mm,top=17mm,bottom=25mm,
inner=17mm,outer=17mm,]{geometry}

\usepackage{cmap} 
\usepackage[T1,T2A]{fontenc}
\usepackage[utf8]{inputenc}
\usepackage[russian,english]{babel}
\usepackage{color}
\usepackage{graphicx}
\usepackage{dcolumn}
\usepackage{bm} 
\usepackage[unicode=true,colorlinks=true,linkcolor=magenta, urlcolor=blue, citecolor = blue,breaklinks]{hyperref}
\usepackage{multirow}
\usepackage{url}
\usepackage{breakurl}
\DeclareGraphicsExtensions{.eps}

\newcount\issue
\newcount\Vol
\newcount\numb
\usepackage{fancyhdr} 
\def\Vol{\textbf{78}}
\def\numb{x}
\begin{document}

\title{Selection of gamma events from IACT images\\with deep learning methods} 

\def\addressa{Irkutsk State University, Applied Physics Institute of Irkutsk State University,\\Russia, 664003, Irkutsk, Gagarin boulevard, 20}
\def\addressb{Lomonosov Moscow State University, Skobeltsyn Institute of Nuclear Physics,\\Russia, 119991, Moscow, Leninskie gory, 1(2)}

\author{\firstname{E.O.}~\surname{Gres}}
\email[E-mail: ]{greseo@mail.ru }
\affiliation{\addressa}
\author{\firstname{A.P.}~\surname{Kryukov}}
\affiliation{\addressb}
\author{\firstname{A.P.}~\surname{Demichev}}
\affiliation{\addressb}
\author{\firstname{J.J.}~\surname{Dubenskaya}}
\affiliation{\addressb}
\author{\firstname{S.P.}~\surname{Polyakov}}
\affiliation{\addressb}
 \author{\firstname{A.A.}~\surname{Vlaskina}}
\affiliation{\addressb}
\author{\firstname{D.P.}~\surname{Zhurov}}
\affiliation{\addressa}


\begin{abstract}
Imaging Atmospheric Cherenkov Telescopes (IACTs) of gamma ray observatory TAIGA detect the Extesnive Air Showers (EASs) originating from the cosmic or gamma rays interactions with the atmosphere. Thereby, telescopes obtain images of the EASs. The ability to segregate gamma rays images from the hadronic cosmic ray background is one of the main features of this type of detectors. However, in actual IACT observations simultaneous observation of the background and the source of gamma ray is needed. This observation mode (called wobbling) modifies images of events, which affects the quality of selection by neural networks.

Thus, in this work, the results of the application of neural networks (NN) for image classification task on Monte Carlo (MC) images of TAIGA-IACTs are presented. The wobbling mode is considered together with the image adaptation for adequate analysis by NNs.  Simultaneously, we explore several neural network structures that classify events both directly from images or through Hillas parameters extracted from images. In addition, by employing NNs, MC simulation data are used to evaluate the quality of the segregation of rare gamma events with the account of all necessary image modifications.

\end{abstract}

\par
\keywords{gamma astronomy, IACT, image recognition, neural networks, wobbling pointing mode.   \\[5pt]}

\maketitle
\thispagestyle{fancy}


\section{Introduction}\label{intro}

Gamma-ray astronomy and astrophysics have been actively developing over the last century and still continue to develop to this day. Research in this area is aimed at studying the processes occurring in supernova explosions, pulsar nebulae, gamma-ray bursts and much more \cite{bib1}. As a result of these phenomena cosmic and gamma radiation with energies of the order of tens and hundreds TeV, which is currently an unreachable task for hadron colliders, are emitted in interstellar medium.

So far in gamma-ray astronomy the registration of high-energy gamma rays is possible only from the Earth due to the small flux of gamma radiation and the large background flux of cosmic radiation. Imaging Atmospheric Cherenkov Telescopes (IACTs) are used for this purpose. Cherenkov radiation from a cascade of particles formed as a result of the interaction of a high-energy particle of primary radiation (atomic nuclei or gamma rays) is recorded by IACT. This cascade in the atmosphere is called an Extensive Air Shower (EAS). Thus, IACT obtains a snapshot of the EAS, which is further processed by the standard method of processing and analysis, called Hillas parameters method\cite{bib2, bib3}. In this method, which is a consequence of themain component method, the registered image from the EAS is parametrized by an ellipse. The characteristics of the ellipse are further used in various cuts and approximations to seperate the desired gamma ray flux from the hadron background. However, today new ways to analyze data from Cherenkov telescopes appeared.

Machine learning methods, including deep learning, are currently actively utilised in all areas of human activity: from an image classification to an analysis and generation of various types of data (images, music, texts, etc). The scientific area, including astrophysics and gamma-ray astronomy, is not an exception. For example, BDT methods are actively used in the MAGIC experiment \cite{bib4}, convolutional neural networks (CNNs) for the HESS installation \cite{bib5, bib6}, as well as in the developing the CTA experiment \cite{bib7, bib8}.

Along with them the TAIGA installation also explores the possibilities of using deep learning methods in the IACT data analysis. TAIGA (Tunka Advanced Instrument for cosmic ray physics and Gamma Astronomy) experiment is a hybrid installation that includes detectors for different components of EASs. One of such detectors is IACT \cite{bib9}. TAIGA facility is located near Lake Baikal in the Tunka Valley, Republic of Buryatia, Russia. Three Cherenkov telescopes have already been installed and are operating on the territory of the facility.

Thus, the aim of this study is to develop deep learning methods for TAIGA–IACT data processing and analisys in order to increase the accuracy and partially automate the processing of TAIGA-IACT data. We note that studies in this area were carried out earlier \cite{bib10, bib11, bib12}, and encouraging results were shown in the classification problem which consists in separating the gamma-ray flux from the hadronic background. In this paper deeper neural networks (NN) analyzing both the images themselves (with CNN) and Hillas parameters (with fully connected NN) are considered. Also the issue of the effect of the wobbling pointing mode, which was not considered in previous works, is taken into account. This mode is important for studying because in real experiment TAIGA-IACTs work in this mode. In addition, the classification task is examined in the case of a strong disbalance of gamma and cosmic radiation fluxes (1:10000).

\section{\label{sec:level1}Wobbling Augmentation for IACT Images}

Wobbling mode is a pointing mode when the currently observed gamma radiation source is not in the center of the camera, but at a certain distance from the center. Periodically, the position of the source is shifted to the opposite location \cite{bib13}. The purpose of such a position wobble is to record events from a gamma source at the so-called ON point (true source) simultaneously with an isotropic background from cosmic rays at the OFF point (false source). ON point is a place on the camera where gamma source is located, while OFF point is any other point on the camera. So it is possible to separate the gamma source signal from the background by comparing the fluxes in the true and false position of the gamma source.

In Monte Carlo (MC) simulation wobbling pointing is ignored: there is no wobble offset of the gamma source, its projection is always in the center of the telescope camera, unlike the experimental data. Since MC simulation is an important component in the training of neural networks, before the possibility of using NNs in the experiment, it was necessary to find a solution to transform images to a uniform view and consider how the displacement impacts TAIGA-IACT image recognition.

\begin{figure}[b]
\includegraphics{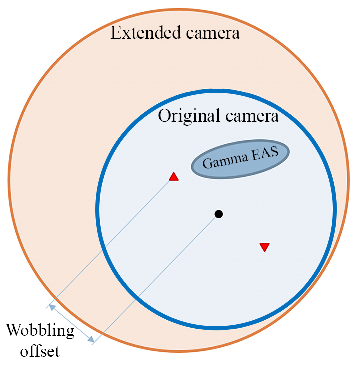}
\caption{\label{fig:fig1} Scheme of the proposed wobbling augmentation necessary to standardize image transformations between experimental data and a neural network’s standard image form.}
\end{figure}

In this paper we propose the following solution for combining simulation and experiment (Fig.~\ref{fig:fig1}). The idea is to build a new camera matrix based on the original camera by adding fake pixels. The center of the expanded matrix must coincide with either the ON or OFF position of the gamma source on the original camera. If the coordinates of the gamma source don't match the coordinates of the pixel, then the coordinates of the pixel closest to the source are assigned as the coordinates of the source. This maneuver is required because IACT image has a discrete (pixel) structure, and the offset is possible up to the pixel size accuracy. For this augmentation it's essential to know the maximum possible wobbling offset, which is true for TAIGA-IACT and is approximately equal to 1.2° (10.1 cm). We have that the simulated images are expanded without displacement (because the source already in the center), while the experimental data is subjected to displacement and expansion of the camera matrix. Therefore, it becomes possible to compare the model and experimental position of the gamma source.

The wobbling mode must also be taken into account when calculating the Hillas parameters. The alignment method is approximately similar to the method described above: the camera coordinate system is shifted to the point under consideration (but without any rewriting of the source coordinates to the nearest pixel), and Hillas parameters are calculated in the new coordinate system. After that the parameters are fed into the neural networks.

\section{Neural Networks for TAIGA-IACT Data Classification}

\subsection{\label{sec:level21}Training samples}

The training set for the neural networks consisted of MC images obtained with the CORSIKA, EAS simulation program \cite{bib14}, and special software of the operation simulation of the TAIGA installation \cite{bib15}. In total, there were 50519 simulated images in the training set, where 25242 were gamma images (gamma ray energy was 1-50 TeV) and 25277 proton images (their energy -- 2-100 TeV). The data was labeled: protons were assigned to class \emph{0}, while gamma events -- to class \emph{1}.

All these events went through the following preprocessing: image cleaning, wobbling data augmentation, image transformation to a square grid and logarithmic scaling. During the cleaning process, pixels with an amplitude less than 7 photoelectrons were set to zero, and after events with a number of non-zero pixels lesser than 5 were removed from the set. This cleaning ensured that there are no noise pixels in the image, and the events themselves in the training set are not on the edge of the telescope's sensitivity. The image transformation to a square form was carried out using the axial method described in \cite{bib16}, because CNNs don't know how to work with hexagonal images. Logarithmic scaling brings the value of pixel amplitudes to the range [0,1], which improved the training of NNs. The scaling happened as follows:

\[ x_{scaled, i} =  \frac{\mathrm{lg}(x_i+1)}{M_{scaled}}, \]
where  \(x_{scaled}\) и \emph{x} are amplitudes of \emph{i}-th pixel after and before scaling, \(M_{scaled}\) is the maximum value among all \(\mathrm{lg}(x_i+1)\) in training samples.

It's worth noting that for Hillas parameters calculation only two steps of preprocessing are required: image cleaning and wobbling data augmentation. Six Hillas parameters were taken into account: \emph{Distance}, \emph{Length}, \emph{Width}, \emph{Alpha}, \emph{Conc} and \emph{lg(Size)}. The first four parameters are parameters characterizing the location of ellipse on the image, while \emph{Conc} and \emph{lg(Size)} are characterizing image brightness\cite{bib2, bib3}. These parameters are the basic set in standard image analysis, and their distribution depends on the type of primary particle, for example, the image from the gamma ray has a more ellipsoidal shape.

\subsection{\label{sec:level22}Considered NN architectures}

\begin{figure*}
\includegraphics{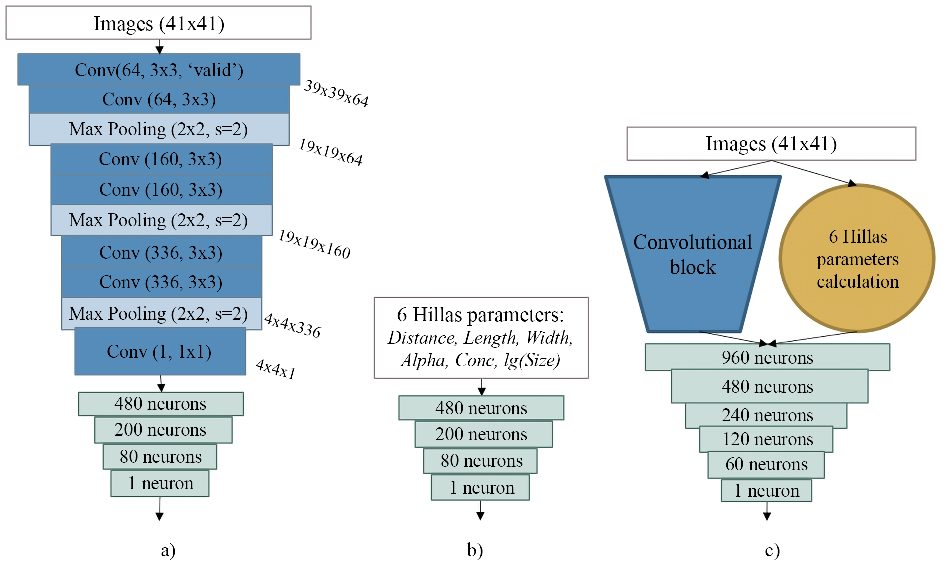}
\caption{\label{fig:fig2}Architecture of neural networks considered in the work in the task of classification of IACT images with wobbling augmentation: a) Convolutional neural network predicting directly from IACT images; b) Fully connected neural network, predicting images' class from Hillas parameters calculated previously from the images; c) Combined convolutional neural network predicting from both images and Hillas parameters.}
\end{figure*}

The classification quality was studied on classificators of the following architectures: convolutional neural network of linear type (Fig.~\ref{fig:fig2}, a); the fully connected NN, where we replaced the block with convolutional layers with the Hillas parameters (\emph{Distance}, \emph{Length}, \emph{Width}, \emph{Alpha}, \emph{Conc} and \emph{lg(Size)}) (Fig.~\ref{fig:fig2}, b), since the parameters are already some essential features of the image, which in the CNN network itself is learning to extract. The combination of these two classification approaches was also considered in the hope that this would help the combined neural network to better generalize and classify events (Fig.~\ref{fig:fig2}, c).

The programming of these networks was carried out using Python's libraries \emph{Tensorflow} and \emph{Keras} \cite{bib17}. The initiation function of neural networks weight coefficients was Xavier uniform initializer. The \emph{ReLU} function was used as the activation function between the inner layers, and the logistic function was applied on the last layer. The loss function was defined as \emph{binary cross entropy}. As a result neural networks output a gamma score for each event, which can be interpreted as a probability measure that the event belongs to a certain class -- gamma class which is \emph{1} or proton class which is \emph{0}.

\subsection{\label{sec:level23}Validation results}

Validation of the NNs training took place on 25283 labeled MC events (12641 gammas and 12642 protons in it), which have been preprocessed in the same way as the training set. The validation set was considered from the two points: ON and OFF points. Events at the ON point, as mentioned above, are only subjected to an expansion of the camera matrix by the wobbling offset distance (10.1 cm) without displacement because we have an MC events. For the OFF point the events were moved to any other point on the camera by 10.1 cm with the camera expansion.

\begin{figure*}
\includegraphics{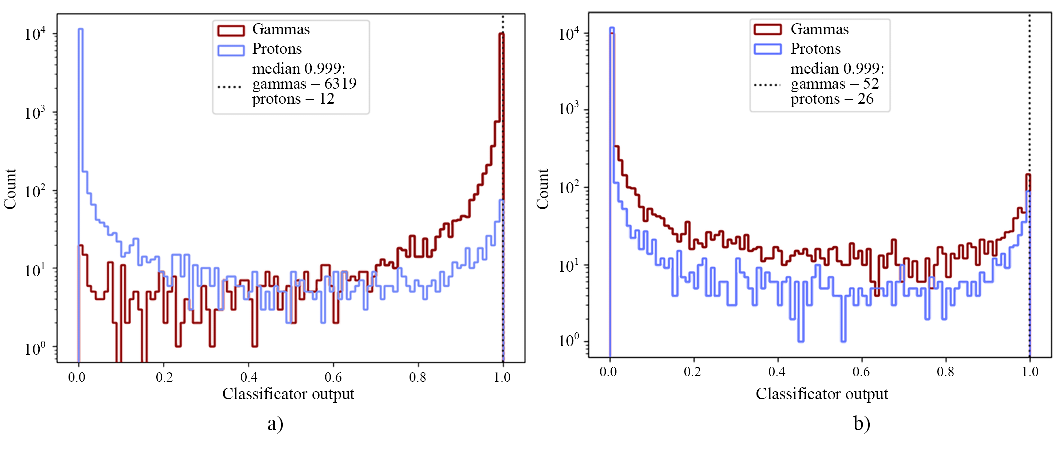}
\caption{\label{fig:fig3} Distributions of NN predictions for gamma and proton events in case of ON (a) and OFF (b) pointing mode. The results of convolutional neural network analyzing only images are presented. Distributions for other NNs differ little from that shown}
\end{figure*}

As the result it was obtained that the area under the ROC curve for each network under consideration was approximately 0.994-0.995 at the ON point and 0.75-0.77 at the OFF point. The distributions of classificator predictions for gamma events and protons are presented on Fig.~\ref{fig:fig3} for CNN (its structure showed in Fig.~\ref{fig:fig2}, a). Other NNs' distributions differ little from that shown. The graphs clearly show that the neural network understands that most gamma events always have a source in the center of the image, which is why gamma events are classified as hadron events at any offset (Fig.~\ref{fig:fig3}, b).

The Table~\ref{tab:table1} shows quantitative classification estimations for each neural network at the ON and OFF points. The threshold of class separation was determined in such a way that after classification about 50 percent of the gamma events were left in the set. The background suppression coefficient \emph{B}, defined as the ratio of past hadrons before and after classification, was used for assessment. The \emph{Q}-factor was also considered, which can be defined as the ratio of the signal significance over the background \cite{bib18} before and after classification:

\begin{equation*}
Q = \frac{S_{after}}{S_{before}}\;,
\end{equation*}
\begin{equation}
S_{after} = \sqrt{2\left(\left(N_{gg}+N_{hg}\right)ln\left(1+\frac{N_{gg}}{N_{hg}} \right) - N_{gg}\right)}\;,
\label{eq:two}
\end{equation}
\begin{equation}
S_{before} = \sqrt{2\left(\left(N_{g}+N_{h}\right)ln\left(1+\frac{N_{g}}{N_{h}} \right) - N_{g}\right)}\;,
\label{eq:three}
\end{equation}
where \(N_{gg}\) and \(N_{hg}\) are gamma and hadron events respectively selected by NN as gamma events, \(N_{g}\) and \(N_{h}\) are gamma and hadron events in training set before classification.

\begin{table*}[htbp]
\renewcommand{\arraystretch}{1.25}
\renewcommand{\tabcolsep}{3pt}
\begin{center}\caption{Estimation of classification results by different neural networks on MC validation set}
\begin{tabular}{|c|c|c|c|c|c|c|c|c|c|c|}\hline
NN model & Classification threshold P & \(N_{total}\)  & \(N_{g}\) & \(N_{h}\) & \(N_{gg}\) & \(N_{hg}\) & \(S_{before}\) & \(S_{after}\) & \emph{Q} & \emph{B} \\ \hline
CNN (ON) & \multirow{2}{*}{0.9985}  & \multirow{6}{*}{25283}  & \multirow{6}{*}{12641} & \multirow{6}{*}{12642} & 6320 & 12 & \multirow{6}{*}{98.82} & 258.35 & 2.61 & 1053.5 \\ \cline{1-1} \cline{6-7} \cline{9-11}
CNN (OFF) & {}                                & {} & {}& {}                                                                                              & 52    & 26 & {}                                 & 8.21 & 0.08 & 486.2 \\ \cline{1-2} \cline{6-7} \cline{9-11}
Fully connected NN (ON)   & \multirow{2}{*}{0.993}   & {} & {}& {}                                                                                              & 6320 & 17 & {}                                 & 249.80 & 2.53 & 743.6 \\ \cline{1-1} \cline{6-7} \cline{9-11} 
Fully connected NN (OFF)  & {}                                & {} & {}& {}                                                                                               & 60    & 17 & {}                                 & 10.61 & 0.11 & 743.6 \\ \cline{1-2} \cline{6-7} \cline{9-11}
Combined CNN (ON)  & \multirow{2}{*}{0.998} & {} & {}& {}                                                                                        & 6320 & 14 & {}                                 & 254.60 & 2.58 & 903.0 \\ \cline{1-1} \cline{6-7} \cline{9-11}
Combined CNN (OFF) & {}                        & {} & {}& {}                                                                                               & 31    & 6 & {}                                  & 8.52 & 0.09 & 2107.0 \\ \hline
\end{tabular}\label{tab:table1}
\end{center}
\end{table*}

It can be seen that results obtained confirm the conclusion that it is necessary to implement data modification for wobbling pointing mode in IACT image analysis with deep learning methods. A small offset greatly worsens the quality of image recognition by neural networks. When examining the results at the ON point for various classificators it might be noticed that \emph{Q}-factor for each NNs doesn't differ much from each other and coefficient \emph{B} is the largest for CNNs. When OFF-pointing, the combined CNN recognized better the absence of a signal. We have that CNN and the combined neural network select events slightly better than the one based on the analysis of Hillas parameters only.

\subsection{\label{sec:level24}Test results}

In the experiment, at least for the Crab Nebula, there is an disbalance of fluxes, approximately 1 gamma event and 10000 hadron events per hour. To consider this desired case a test set was compiled, which contained 92606 experimental hadrons (already with wobbling offset) and 9 MC gamma events, specially shifted at the modeling stage by 10 cm for testing neural networks. To get closer to the experimental data processing, where the number of gamma and hadron events is unknown, eleven OFF points on the camera were selected together with one ON point. The true and false sources were located relative to each other every 30 degrees on a circle with a radius of the wobbling offset. Based on the hadrons passed through classificators at 11 OFF points the mean and standard deviation of the hadron background is calculated in experimental data processing.

Table~\ref{tab:table2} presents the number of events for 11 OFF and one ON point sources obtained as a result of NNs testing. At OFF points, no gamma event exceeded the classification threshold. It is easy to notice that the coefficient \emph{B} is, according to the table, approximately 840-1100, which corresponds to the values obtained during validation. However, for such a strong disbalance of classes, the signal from the gamma source does not even exceed \(1\sigma\) for any neural network (the best result belongs to CNN). Using statistical estimation mentioned in Section~\ref{sec:level23}, significance of the gamma signal \(S_{after}\) and \(S_{before}\) (Eq.~(\ref{eq:two}) and (\ref{eq:three})) are 0.49-0.66 and 0.03 respectively, leading to Q-factor being 16.61-22.33 (the highest Q-factor value belongs to fully connected NN, based on Hillas parameters).

\begin{table*}[htbp]
\renewcommand{\arraystretch}{1.25}
\renewcommand{\tabcolsep}{3pt}
\begin{center}\caption{Estimation of classification results by different neural networks on test set consisting from experimental hadron events and MC gamma events }
\begin{tabular}{|c|c|c|c|c|c|} \hline
\multirow{2}{*}{NN model} & \multirow{2}{*}{Classification threshold P} & \multicolumn{2}{|c|}{ON} & \multicolumn{2}{|c|}{OFF} \\ \cline{3-6}
{} & {} & True gamma & Fake gamma (hadrons) & Mean & \(\sigma\) \\ \hline
CNN & 0.9985 & 5 & 106 & 102.27 & 5.27      \\ \hline 
Fully connected NN & 0.993  & 7 & 88 & 110.18 & 13.43       \\ \hline 
Combined CNN  & 0.993  & 5 & 83 & 98.00 & 10.50 \\ \hline 
\end{tabular}\label{tab:table2}
\end{center}
\end{table*}

We would like to note that in previous works devoted to the classification of TAIGA-IACT data by neural networks class disbalance -- up to 1:1000 -- was also considered \cite{bib12}. In that work, model data were considered with more lenient image cleaning criteria (cleaning threshold was 5 phe) and without wobbling augmentation. Compared to the previous results there is a noticeable improvement in the hadron background suppression, the coefficient of which reaches the value of 1000 (in the previous case - up to 200).

\section{Conclusion}

Within the frames of this work the issue of the effect of the wobbling pointing mode, which leads to differences in MC and experimental TAIGA-IACT data, on the quality of classification by various neural networks was studied. The results clearly show that image adaptation to Woobling mode is essential in IACT image recognition. In further work on deep learning methods it's necessary to take into account the effect of wobbling on the quality of classification.

At the same time, it was found out that there are no major differences in the classification results of the CNN (from images), fully connected NN (from Hillas parameters) and the combined CNN (from images and Hillas parameters), so in the future only one way of neural network analysis of input TAIGA-IACT data can be considered.

The study of the application of deep learning methods to TAIGA-IACT data analysis has shown that neural networks considered in this paper can strongly suppress the hadron background (hadron suppression value reaches up to 1000), yet the signal still does not exceed the level of \(1\sigma\) for gamma and hadron event ratio ``1:10000''. As a result, a substantial improvement of this method is still required for the application of deep learning methods in the experimental data processing to gain a good level of analysis. Improvements may include expanding the training set (by number and energy range of events), fine adjustments in architacture and hyperparameters of the neural networks, different data preprocessing (like cleaning) and others. On the basis of the foregoing we would like to note that the refinement of deep learning methods for TAIGA-IACT data processing and analysis will be continued.

\begin{acknowledgments}
The authors would like to express their gratitude to the staff of TAIGA collaboration for their support and recommendations during this work. Also the work was done using the data of UNU ``Astrophysical Complex of MSU-ISU'' (agreement EB-075-15-2021-675).
\end{acknowledgments}

\section*{FUNDING}
This work was supported by RSF, grant no.22-21-00442.

\end{document}